\documentclass[sigconf]{acmart}
\usepackage{algorithmic}
\usepackage{algorithm}
\usepackage{bm}
\usepackage{mathtools}

\usepackage{amsmath}
\usepackage{array}
\usepackage{tabulary}
\usepackage{arydshln}

\usepackage{enumitem}

\usepackage{float}

\usepackage{graphicx}
\usepackage{subfigure}
\usepackage{listings}
\usepackage{color}
\definecolor{dkgreen}{rgb}{0,0.6,0}
\definecolor{gray}{rgb}{0.5,0.5,0.5}
\definecolor{mauve}{rgb}{0.58,0,0.82}

\setcopyright{None}
\renewcommand\footnotetextcopyrightpermission[1]{} 
\begin{document}


\title{FinRL-Podracer: High Performance and Scalable Deep Reinforcement Learning for Quantitative Finance}


\author{Zechu Li}
\email{zl2993@columbia.edu}
\affiliation{%
  \institution{Columbia University}
}

\author{Xiao-Yang Liu}
\authornote{Equal contribution.}
\email{xl2427@columbia.edu}
\affiliation{%
  \institution{Columbia University}
}

\author{Jiahao Zheng}
\email{jh.zheng@siat.ac.cn}
\affiliation{%
  \institution{Shenzhen Inst. of Advanced Tech.}
}

\author{Zhaoran Wang}
\email{zhaoranwang@gmail.com}
\affiliation{%
  \institution{Northwestern University}
}

\author{Anwar Walid}
\authornote{A. Walid finished this project at Bell labs, before joining Amazon.}
\email{anwar.i.walid@gmail.com}
\affiliation{%
  \institution{Amazon \& Columbia University}
}

\author{Jian Guo}
\authornote{Corresponding author.}
\email{guojian@idea.edu.cn}
\affiliation{%
  \institution{IDEA Research}
}

\begin{abstract}

Machine learning techniques are playing more and more important roles in finance market investment. However, finance quantitative modeling with conventional supervised learning approaches has a number of limitations, including the difficulty in defining appropriate labels, lack of consistency in modeling and trading execution, and lack of modeling the dynamic nature of the finance market. The development of deep reinforcement learning techniques is partially addressing these issues. Unfortunately, the steep learning curve and the difficulty in quick modeling and agile development are impeding finance researchers from using deep reinforcement learning in quantitative trading. In this paper, we propose an \textit{RLOps in finance} paradigm and 
present a \textit{FinRL-Podracer} framework to accelerate the development pipeline of deep reinforcement learning (DRL)-driven trading strategy and to improve both trading performance and training efficiency. FinRL-Podracer is a cloud solution that features high performance and high scalability and promises \textit{continuous training}, \textit{continuous integration}, and \textit{continuous delivery} of DRL-driven trading strategies, facilitating a rapid transformation from algorithmic innovations into a profitable trading strategy. First, we propose a generational evolution mechanism with an ensemble strategy to improve the trading performance of a DRL agent, and schedule the training of a DRL algorithm onto a GPU cloud via multi-level mapping. Then, we carry out the training of DRL components with high-performance optimizations on GPUs. Finally, we evaluate the FinRL-Podracer framework for a stock trend prediction task on an NVIDIA DGX SuperPOD cloud. FinRL-Podracer outperforms three popular DRL libraries \textit{Ray RLlib},  \textit{Stable Baseline 3} and \textit{FinRL}, i.e., $12\% \sim 35 \%$ improvements in annual return, $0.1 \sim 0.6$ improvements in Sharpe ratio and $3\times \sim 7\times$ speed-up in training time. We show the high scalability by training a trading agent in $10$ minutes with $80$ A100 GPUs, on NASDAQ-100 constituent stocks with minute-level data over $10$ years.

\end{abstract}

\ccsdesc[500]{Computing methodologies~Machine learning}
\ccsdesc[500]{Computing methodologies~Neural networks}
\ccsdesc[500]{Computing methodologies~Markov decision processes}
\ccsdesc[500]{Computing methodologies~Reinforcement learning}

\keywords{RLOps in finance, deep reinforcement learning, stock trend prediction, scalability, GPU cloud}

\maketitle

 \section{Introduction}





\begin{figure}[t]
\centering
\includegraphics[width=3.3in]{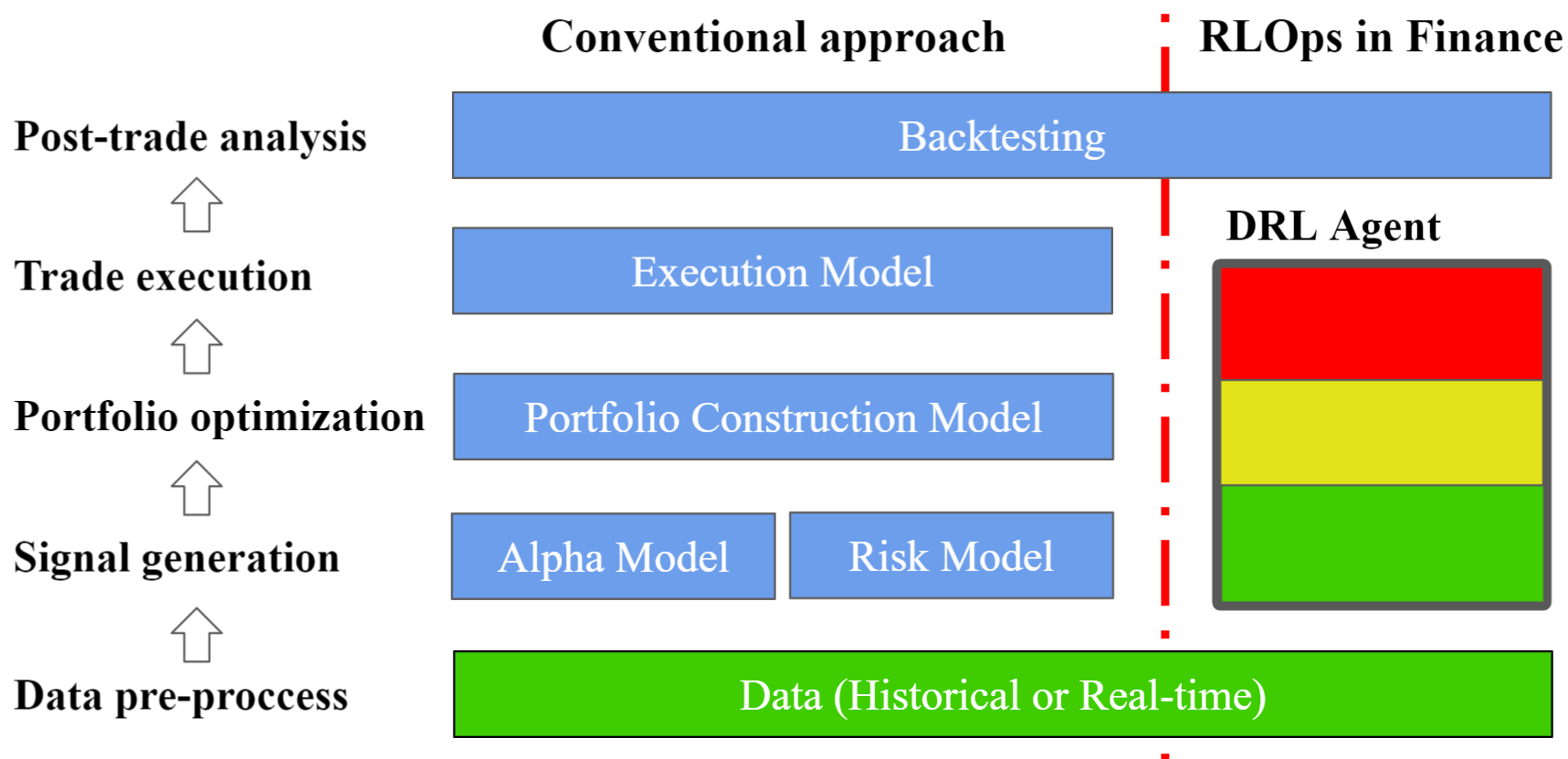}
\caption{Software stack for an algorithmic trading process: conventional approach vs. RLOps in finance.}
\vspace{-0.2in}
\label{fig_structure}
\end{figure}

Algorithmic trading is increasingly deployed in the financial investment process. A conventional supervised learning pipeline consists of five stages \cite{Treleaven2013AlgorithmicTR, Nuti2011AlgorithmicT}, as shown in Fig. \ref{fig_structure} (left), namely data pre-process, modeling and trading signal generation, portfolio optimization, trade execution, and post-trade analysis. Recently, deep reinforcement learning (DRL) \cite{silver2016mastering, silver2017mastering, sutton2018reinforcement} has been recognized as a promising alternative for quantitative finance \cite{Kolm2019ModernPO, Buehler2019DeepH, jpmorgan, McKinsey}, since it has the potential to overcome some important limitations of supervised learning, such as the difficulty in label specification and the gap between modeling, positioning and order execution. We advocate extending the principle of \textit{MLOps} \cite{alla2021mlops} \footnote{MLOps is an ML engineering culture and practice that aims at unifying ML system development (Dev) and ML system operation (Ops).} to the \textit{RLOps in finance} paradigm that implements and automates the continuous training (CT), continuous integration (CI), and continuous delivery (CD) for trading strategies. We argue that such a paradigm has vast profits potential from a broadened horizon and fast speed, which is critical for wider DRL adoption in real-world financial tasks.


The \textit{RLOps in finance} paradigm, as shown in Fig. \ref{fig_structure} (right), integrates middle stages (i.e., modeling and trading signal generation, portfolio optimization, and trade execution) into a DRL agent. Such a paradigm aims to help quantitative traders develop an end-to-end trading strategy with a high degree of automation, which removes the latency between stages and results in a compact software stack. The major benefit is that it can explore the vast potential profits behind the large-scale financial data, exceeding the capacity of human traders; thus, the trading horizon is lifted into a potentially new dimension. Also, it allows traders to continuously update trading strategies, which equips traders with an edge in a highly volatile market. However, the large-scale financial data and fast iteration of trading strategies bring imperative challenges in terms of computing power.

Existing works are not satisfactory with respect to the usage of large-scale financial data and the efficiency of agent training. For DRL strategy design, existing works studied algorithmic trading on a daily time-frame \cite{liu2020finrl, liu2021finrl, yang2020stock, xiong2018practical, zhang2019stock, 8851831, Buehler2019DeepH} or hourly time-frame \cite{jiang2017portfolio}, which is hard to fully 
capture the dynamics of a highly volatile market. For DRL library/package development, existing works may not be able to meet the intensive computing requirement of relatively high frequency trading tasks, large-scale financial data processing and tick-level trade execution. We evaluate the training time of three popular DRL libraries, FinRL \cite{liu2020finrl, liu2021finrl}, RLlib \cite{liang2017rllib} and Stable Baseline3 \cite{stablebaseline3}, on NASDAQ-100 constituent stocks with minute-level data (described in Section \ref{metrics}) in Table \ref{complexity_comparison}, which shows that it is difficult for them to effectively train a profitable trading agent in a short cycle time.

In recent years, distributed DRL frameworks and massively parallel simulations have been recognized as the critical software development for the RLOps paradigm \cite{liang2017rllib, hessel2021podracer, Liang2018GPUAcceleratedRS, makoviychuk2021isaac}. It is promising to utilize extensive computing resources, e.g., a GPU cloud, to accelerate the development of trading strategies, including both financial simulation and model training. Therefore, we investigate a candidate solution on a GPU cloud, an NVIDIA DGX SuperPOD cloud \cite{NVIDIA_SupPod2020} that is the most powerful AI infrastructure for enterprise deployments.

In this paper, we propose a \textit{FinRL-Podracer} framework as a high-performance and scalable solution for \textit{RLOps in finance}. At a high level, FinRL-Podracer schedules the training process through a multi-level mapping and employs a generational evolution mechanism with an ensemble strategy. Such a design guarantees scalability on a cloud platform. At a low level, FinRL-Podracer realizes hardware-oriented optimizations, including parallelism encapsulation, GPU acceleration, and storage optimization, thus achieving high performance. As a result, FinRL-Podracer can effectively exploit the supercomputing resources of a GPU cloud, which provides an opportunity to automatically design DRL trading strategies with fast and flexible development, deployment and production.

Our contributions can be summarized as follows
\begin{itemize}
    \item We propose a \textit{FinRL-Podracer} framework built on two previous projects, FinRL \cite{liu2021finrl, liu2020finrl} and ElegantRL \cite{erl} \footnote{ElegantRL is a scalable and elastic deep reinforcement learning library. It supports general robotic and game playing applications.}, to initiate a paradigm shift from conventional supervised learning approaches to \textit{RLOps in finance}. 
    \item FinRL-Podracer employs a generational evolution mechanism during the training of DRL agents and provides high-performance optimizations for financial tasks. 
    \item We show the training efficiency by obtaining a trading agent in $10$ minutes on an NVIDIA DGX SuperPOD cloud \cite{NVIDIA_SupPod2020} with $80$ A100 GPUs, for a stock trend prediction task on NASDAQ-100 constituent stocks with minute-level data over $10$ years. We evaluate the trained agent for one year and show its trading performance outperforms three DRL libraries, \textit{FinRL}  \cite{liu2020finrl, liu2021finrl}, \textit{Ray RLlib} \cite{liang2017rllib} and \textit{Stable Baseline3} \cite{stable-baselines}, i.e., $12\% \sim 35 \%$ improvements in annual return, $0.1 \sim 0.6$ improvements in Sharpe ratio and $3\times \sim 7\times$ speed-up in training time. 
\end{itemize}

The remainder of this paper is organized as follows. Section 2 describes related works. Section 3 models a typical stock trend prediction task as a Markov Decision Process. In Section 4, we present the FinRL-Podracer framework and describe its evolution and training layers, respectively. In Section 5, we describe the experimental setup for the trading task and present experimental results. We conclude this paper and discuss future directions in Section 6.

\begin{table}[t]
		\centering
		\begin{tabular}{|l|c|c|c|}\hline
			  & Sharpe ratio & Max dropdown & Training time \\
			\hline
			RLlib \cite{liang2017rllib} & 1.67 & \textbf{-23.248\%} & \textbf{110 min} \\
			SB3 \cite{stable-baselines} & \textbf{1.82} & -23.750\% & 150 min \\
			FinRL \cite{yang2020stock} & 1.35 & -27.267\% & 345 min\\
			QQQ & 1.25 & -28.559\% & --\\
			\hline
		\end{tabular}
		\caption{Evaluations of existing DRL libraries on an NVIDIA DGX A100 server \cite{choquette2021nvidia}. We evaluate on NASDAQ-100 constituent stocks with minute-level data by training from 01/01/2009 to 05/12/2019 and backtesting from 05/13/2019 to 05/26/2021. Invesco QQQ ETF is a market benchmark.}
        \label{complexity_comparison}
		\vspace{-0.15in}
\end{table}

\section{Related Works}

This section summarizes related works from two aspects: DRL applications in quantitative finance and the MLOps development.  

\subsection{DRL in Finance}

With the successes of DRL in game playing, e.g., Atari games \cite{mnih2015atari} and GO games \cite{silver2016go}, more and more finance researchers show their interests in this area, and they have done some early attempts to applying DRL in quantitative finance investment. In this paper, we take the stock trend prediction (STP) task as an example to introduce existing works and show great potentials of DRL in finance area. 

Stock trend prediction task is often considered a challenging application of machine learning in finance due to its noisy and volatile nature. Traditionally, the STP task is formulated as a supervised learning problem, where features of stocks are extracted from a past time window of technical indices, fundamental data and alternative data \cite{Chen2020QuantifyingEA}, and labels are usually extracted from a future time window of concerned criteria such as rise/fall, returns, excess returns or Sharpe ratios. Recently, deep reinforcement learning has been applied to solving STP tasks. Zhang \textit{et al}. \cite{zhang2019stock} constructed a trading agent using three DRL algorithms, DQN, PG, and A2C, for both discrete and continuous action spaces. Yang \textit{et al}. \cite{yang2020stock} used an ensemble strategy to integrate different DRL algorithms, A2C, DDPG, and PPO based on the Sharpe ratio. They applied the idea of a rolling window, where the best algorithm is picked to trade in the following period. Recently, many researchers provide more DRL solutions for STP tasks \cite{dang2019stock, Chen2019ApplicationOD, li2019bull}. However, most existing works are based on several assumptions, which limits the practicality. For example, the backtesting has no impacts on the market; there are almost no other complex actions besides buying, holding, and selling; only one stock type is supported for each agent.

\subsection{Principle of MLOps}

Recently, Google trends put Machine Learning Operations (MLOps) as one of the most promisingly increasing trends \cite{mlopstrend}. MLOps is a practice in developing and operating large-scale machine learning systems, which facilitates the transformation of machine learning models from development to production \cite{mlopsgoogle, mlopsnvidia}. In essence, MLOps entails cloud computing power to integrate and automate a standard machine learning pipeline: 1) data pre-processing; 2) feature engineering; 3) continuous ML model training; 4) continuous ML model deployment; 5) output production, thus building applications that enable developers with limited machine-learning expertise to train high-quality models specific to their domain or data \cite{mlopssustainable, mlopscloud}. 

However, the DRL is quite different from conventional machine learning approaches. For example, training data of DRL is not prepared in advance compared with conventional supervised learning but collected through an agent-environment interaction inside the training process. Such a significant difference requires a re-integration of the automated pipeline and a re-scheduling of the cloud computing resources with respect to the conventional MLOps principle. Therefore, we advocate extending the principle of MLOps to the \textit{RLOps in finance} paradigm to seek an opportunity for the wider DRL adoption in production-level financial services.

\begin{figure*}[t]
\centering
\includegraphics[width=7in]{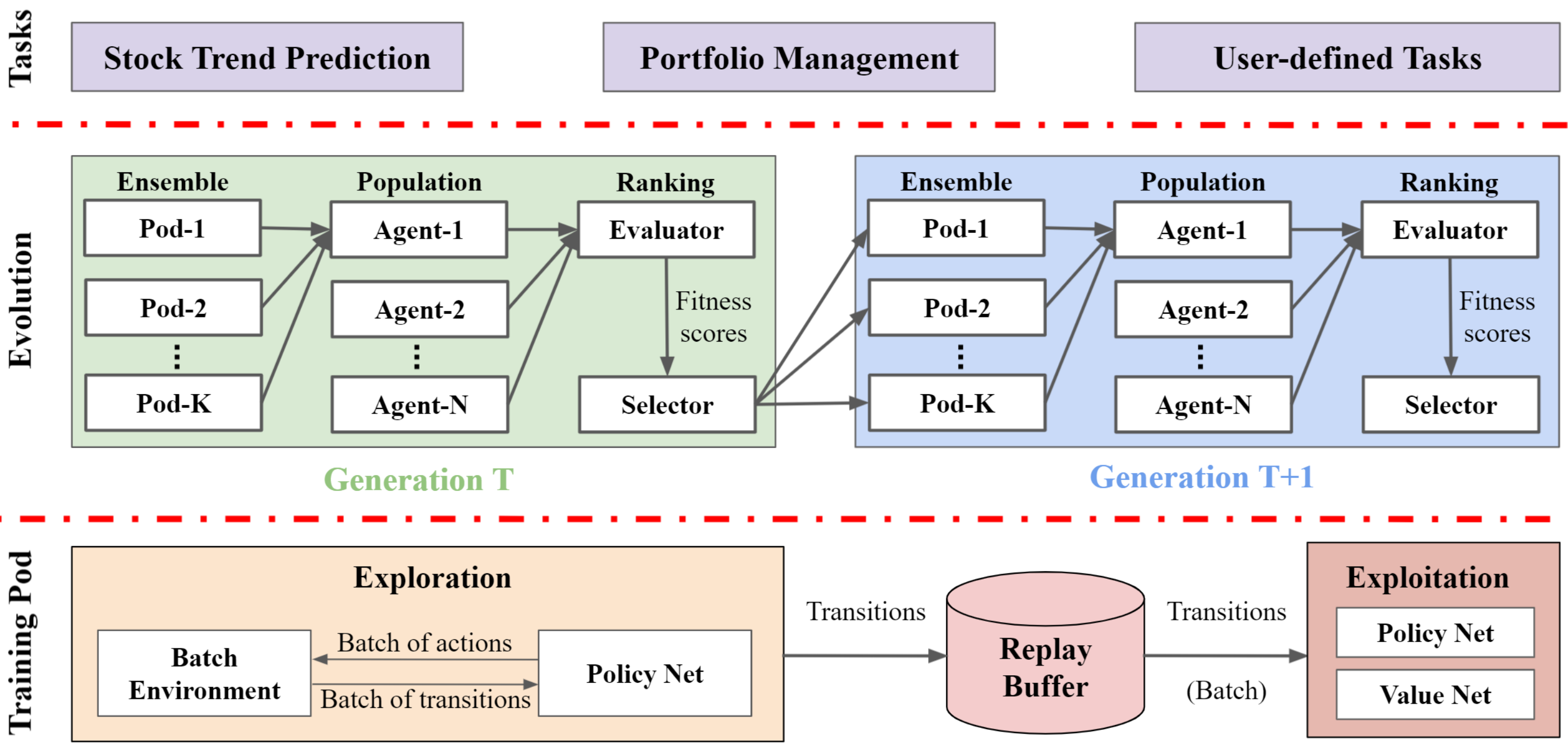}
\caption{Overview of \textit{FinRL-Podracer} that has three layers: trading task layer, evolution layer and training layer.}
\label{fig_framework}
\end{figure*}

\section{Stock Trend Prediction Task}
\label{problem_formulation}

We describe the problem formulation of a typical financial task, stock trend prediction, which locates at the task layer of Fig. \ref{fig_framework}. Our setup follows a similar setting in \cite{liu2021finrl, yang2020stock}.


A stock trend prediction task is modeled as a Markov Decision Process (MDP): given state $s_t \in \mathcal{S}$ at time $t$, an agent takes an action $a_t \in \mathcal{A}$ according to policy $\pi_{\theta}(s_t)$, transitions to the next state $s_{t+1}$ and receives an immediate reward $r(s_t, a_t, s_{t+1}) \in \mathbb{R}$. The policy $\pi_\theta(s)$ with parameter $\theta$ is a function that maps a state to an action vector over $n$ stocks. The objective is to find an optimal policy $\pi^*_\theta$ (a policy network parameterized by $\theta$) that maximizes the expected return (the fitness score used in Fig. \ref{fig_framework}) over $T$ times slots
\begin{equation}\label{obj_stock}
    \pi^*_\theta = \underset{\theta}{\text{argmax}}~ J(\pi_\theta),~~\text{where}~~J(\pi_\theta) = \mathbb{E}\left[\sum^{T}_{t=0}\gamma^tr(s_t, a_t, s_{t+1})\right], 
\end{equation}
where $\gamma \in (0,1]$ is a discount factor. 

Then, for the stock trend predictiont task with $n$ stocks, we specify the state space $\mathcal{S}$, action space $\mathcal{A}$, reward function $r(s_t, a_t, s_{t+1})$, and the state transition, as in \cite{liu2021finrl}\cite{yang2020stock}.

\textbf{State space} $\mathcal{S}$ describes an agent's perception of a market environment. We summarize various features that are used by human trader and use them to construct the state space:
\begin{itemize}[leftmargin=*]
    \item Balance $b_t \in \mathbb{R_+}$: the account balance at time $t$.
    \item Shares $\textbf{h}_t \in \mathbb{Z}^n_+$: the number of shares for $n$ stocks at $t$.
    \item Closing price $\textbf{p}_t \in \mathbb{R}^n_+$: the closing prices of $n$ stocks at $t$.
    \item Technical indicators help the agent make decisions. Users can select existing indicators or add new indicators. E.g., Moving Average Convergence Divergence (MACD) $\textbf{M}_t \in \mathbb{R}^n$, Relative Strength Index (RSI) $\textbf{R}_t \in \mathbb{R}^n_+$, Commodity Channel Index (CCI) $\textbf{C}_t \in \mathbb{R}^n_+$, etc. 
\end{itemize}

\textbf{Action space} $\mathcal{A}$ describes the allowed actions an agent can take at states $s_t, t=1, ..., T$. For one stock, action is $a \in \{-k, ..., -1, 0, 1, ..., k\}$, where $k \in \mathbb{Z}$ or $-k \in \mathbb{Z}$ denotes the number of shares to buy or sell, respectively, and $a = 0$ means to hold. Users can set a maximum number of shares $h_{\max}$ for a transaction, i.e., $k \leq h_{\max}$, or set a maximum ratio of capital to allocate on each stock.

\textbf{Reward} $r_t$ for taking action $a_t$ at state $s_t$ and arriving at state $s_{t+1}$. Reward is the incentive for an agent to improve its policy for the sake of getting higher rewards. A relatively simple reward can be defined as the change of the account value, i.e.,
\begin{equation}\label{obj_stock2}
    r_t = (b_{t+1}+\textbf{p}^T_{t+1}\textbf{h}_{t+1}) - (b_t+\textbf{p}^T_t\textbf{h}_t) - c_t, 
\end{equation}
where the first and second terms are the account values at $s_{t+1}$ and $s_t$, and $c_t$ denotes the transaction cost (market friction).

\textbf{Transition} $(s_t, a_t, r_t, s_{t+1})$. Taking  action $a_t$ at state $s_t$, the environment steps forward and arrives at state $s_{t+1}$. A transition involves the change of balance, number of shares, and the stock prices due to the market changes. We split the stocks into three sets: selling set $S$, buying set $B$ and holding set $H$, respectively. The new balance is
\begin{equation}\label{obj_stock3}
    b_{t+1} = b_t + (\textbf{p}_t^S)^T\textbf{k}_t^S - (\textbf{p}_t^B)^T\textbf{k}_t^B, 
\end{equation}
where $\textbf{p}^S \in \mathbb{R}^{n}$ and $\textbf{k}^S \in \mathbb{R}^{n}$ are the vectors of prices and number of selling shares for the selling stocks, and $\textbf{p}^B \in \mathbb{R}^{n}$ and $\textbf{k}^B \in \mathbb{R}^{n}$ for the buying stocks. The number of shares becomes
\begin{equation}\label{obj_stock4}
    \textbf{h}_{t+1} = \textbf{h}_t - \textbf{k}_t^S + \textbf{k}_t^B \geq \bm{0}. 
\end{equation}

The supercomputing power is necessary to achieve the massively parallel simulations for an STP task. During the training, a DRL agent keeps observing and trading on the historical market data to sample trajectories (one trajectory is a series of transitions). However, the historical data has to be significantly large in order to provide a broadened horizon. For example, the historical data could scale up in two dimensions: the \textbf{data volume} and \textbf{data type} \cite{jpmorgan}:
\begin{itemize}[leftmargin=*]
    \item The data volume varies with respect to:
        \begin{itemize}
            \item \textbf{the length of data period}: from several months up to more than ten years.
            \item \textbf{the time granularity}: from daily-level to minute-level, second-level or microsecond-level.
            \item \textbf{the number of stocks}: from thirty (Dow 30) to hundreds (NASDAQ 100 or S\&P 500), or even covers the whole market.
          \end{itemize}
    \item The data type varies with respect to:
        \begin{itemize}
            \item \textbf{the raw market data} includes data points of open-high-low-close-volume (OHLCV) for each stock, which provides a direct understanding of a stock's market performance.
            \item \textbf{the alternative data} usually refers to  the large-scale collection of both structured and unstructured data, e.g., market news, academic graph data \cite{Chen2020QuantifyingEA}, credit card transactions and GPS traffic. The agent could employ different encoders to analyze the insights of investment techniques provided by the alternative data.
            \item \textbf{the indexes and labels} could be directly given as a kind of powerful technical indicator, which helps the agent make decisions. 
          \end{itemize}
\end{itemize}

In practice, the market simulation, alternative data processing and index analyzing are computationally expensive, therefore, a cloud-level solution is critical to a fast iteration of a trading strategy.

\section{FinRL-Podracer Framework}

We propose a \textit{FinRL-Podracer} framework to utilize the supercomputing power of a GPU cloud for training DRL-driven trading strategies. We first present an overview of FinRL-Podracer and then describe its layered architecture.

\subsection{Overview}

Based on the experiments in Table \ref{complexity_comparison}, we found that existing DRL libraries/packages \cite{liang2017rllib, stable-baselines, liu2020finrl, liu2021finrl} have three major issues that restrict the trading performance and training efficiency: 
\begin{itemize}[leftmargin=*]
    \item There is no criteria to determine \textbf{overfitting or underfitting of models (trading strategies)} during the training process. It is critical to overcome underfitting by utilizing more computing power and avoid overfitting that wastes computing power, while both cases would lead to suboptimal models.
    \item The training process of a trading strategy is \textbf{sensitive to hyper-parameters}, which may result in unstable trading performance in backtesting and trading. However, it is tedious for human traders to search for a good combination of hyper-parameters, and thus an automatic hyper-parameter search is favored.
    \item \textbf{Computing power is critical} to effectively explore and exploit large-scale financial data. Sufficient exploration guarantees a good trading performance, and then smart exploitation results in good training efficiency. A strong computing power helps achieve a balance between exploration and exploitation.  
\end{itemize}


Therefore, we provide a high performance and scalable solution on a GPU cloud, \textit{FinRL-Podracer}, to develop a profitable DRL-driven trading strategy within a small time window. To fully utilize a GPU cloud, say an NVIDIA DGX SuperPod cloud \cite{NVIDIA_SupPod2020}, we organize FinRL-Podracer into a three-layer architecture in Fig. \ref{fig_framework}, a trading task layer on the top, an evolution layer in the middle and a training layer at the bottom. 

In the evolution layer, we employ a generational evolution mechanism with the ensemble strategy and address the issues of overfitting and hyper-parameter sensitivity through the synergy of an \textit{evaluator} and a \textit{selector}. The evaluator computes the fitness scores $J(\pi_\theta)$ in (\ref{obj_stock}) of a population of $N$ agents and mitigates the performance collapse caused by overfitting. The hyper-parameter search is automatically performed via \textit{agent evolution}, where the selector uses the fitness scores to guide the search direction. An effective cloud-level evolution requires a high-quality and scalable scheduling, therefore we schedule a population of parallel agents through a multi-level mapping.

In the training layer, we realize high-performance GPU-oriented optimizations of a decomposable DRL training pipeline. We locally optimize each component (a container within a pod), namely explorer, replay buffer, and learner, through parallelism encapsulation, GPU acceleration, efficient parameter fusion, and storage optimization. Thus, we maximize the hardware usage and minimize the communication overhead, which allows each component to be efficiently executed on a GPU cloud.

Such an evolution-and-training workflow pipelines the development of a trading strategy on a GPU cloud. It enjoys great performance and scalability, which promotes fast and flexible development, deployment and production of profitable DRL trading strategies.

\subsection{Scalable Evolution Layer}
\label{orchestration}

FinRL-Podracer exploits a \textit{generational evolution} mechanism with an ensemble strategy to coordinate the parallel agents and to automatically search the best hyper-parameters. For each generation, it is composed of \textbf{model ensemble} and \textbf{population ranking}, as shown in the middle layer of Fig. \ref{fig_framework}. At present, we utilize an \textit{evaluator} and a \textit{selector} to schedule the agent evolution, where more modules can be incorporated, e.g., a monitor, an allocator, etc.

The evaluator evaluates agents and provides their fitness scores as the metric for the future ranking, as shown in the ranking stage of Fig. \ref{fig_framework}. From our observations, it is difficult for users to use existing libraries \cite{liang2017rllib, stable-baselines, liu2020finrl, liu2021finrl} to train a profitable trading strategy because the overfitting agent may be treated as the best agent as the training process moves forward. When the dataset scales up, we need to increase the training time/steps to fully explore the large-scale data, making it harder to set appropriate stop criteria, and the resulting agent may hardly be the best one. The evaluator effectively mitigates the performance collapse brought by overfitting: in the course of the training, it evaluates the agent at each iteration, outputs a fitness score, and keeps track of the best agent so far; when the fitness score in (\ref{obj_stock}) drops, the evaluator would stop the training process using the early stopping mechanism and output the best agent as the final agent.

The selector acts as a central controller to perform the selection strategy as in a genetic algorithm (GA) \cite{Mitchell1996AnIT}. GA is an optimization algorithm inspired by natural evolution: at every generation, a population of N agents is trained, and the evaluator calculates their fitness scores in (\ref{obj_stock}) based on an objective function; then the selector redistributes the agents with the highest scores to form a new population for the next generation. Since the agents are parallel and replicable, the concept of natural selection scales up well on a GPU cloud. As shown in the evolution layer of Fig. \ref{fig_framework}, there are $N$ agents with different hyper-parameters in a population. The synergy of the evaluator and selector enables FinRL-Podracer to naturally select the best agent for the future generation and eliminates the potential negative impact from poorly evolved agents, which effectively improves the stability and efficiency of the training. 

FinRL-Podracer achieves the ensemble training of an agent by concurrently running K pods (training processes) in parallel and fusing the trained models from K pods at each epoch. All parallel pods for each agent are initialized with same hyper-parameters but different random seeds. Such a design, as shown in the ensemble stage of the evolution layer in Fig. \ref{fig_framework}, guarantees randomness and stabilizes the learning process of the agent. The experiment results in Section \ref{evaluation} will perform an ablation study of the improvement brought by the generational evolution mechanism.

\definecolor{arylideyellow}{rgb}{0.91, 0.84, 0.42}

\begin{figure}
\centering
\includegraphics[width=3.3in]{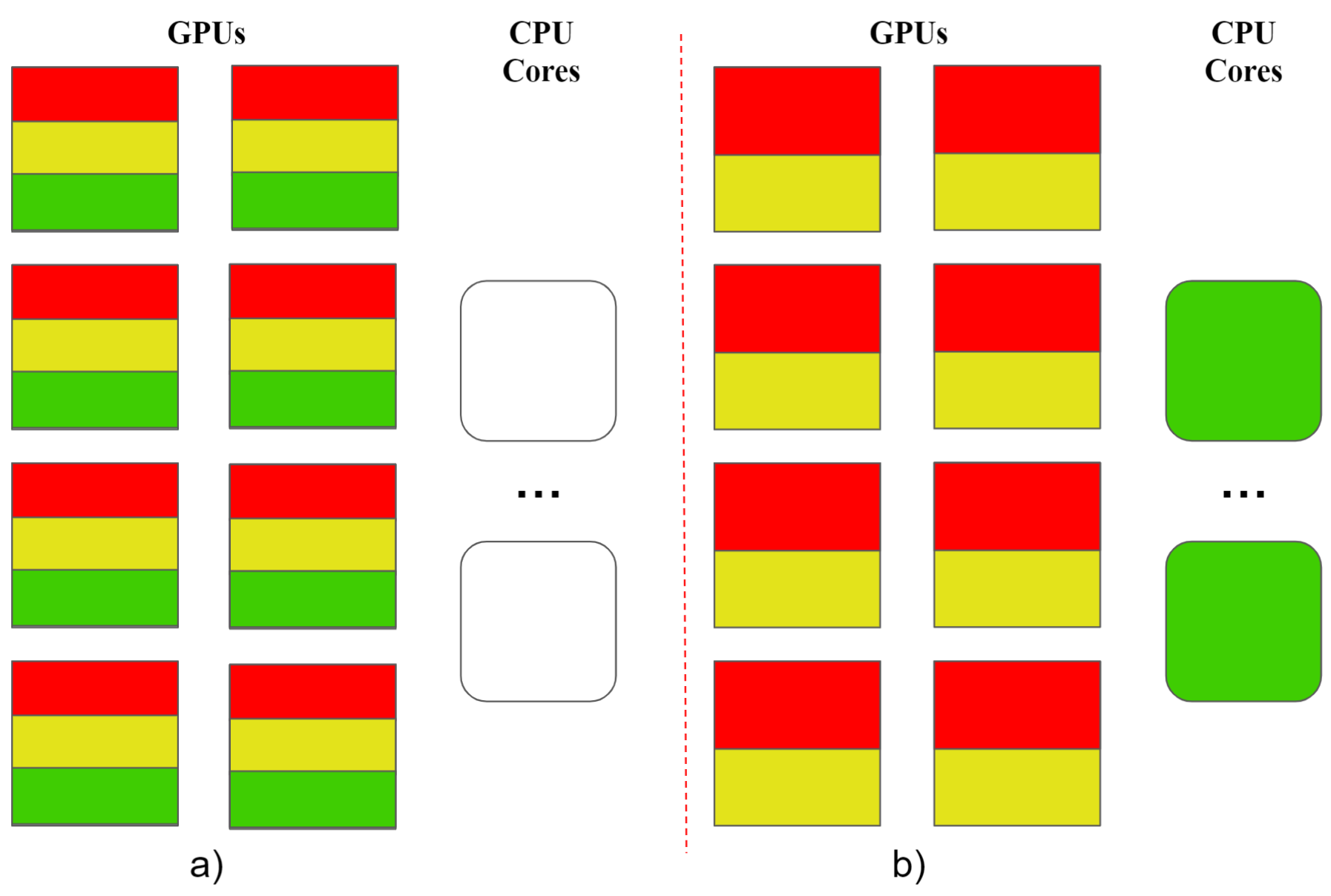}
\caption{Two implementations of a training process: a) the environment simulation (\textcolor{green}{green}), action inference (\textcolor{arylideyellow}{yellow}), and model update (\textcolor{magenta}{red}) are all located on GPUs. b) the environment simulation is executed on CPUs, and action inference and model update are on GPUs.  An NVIDIA DGX-A100 server \cite{NVIDIA_SupPod2020}\cite{choquette2021nvidia} contains $8$ A100 GPUs.}
\label{fig_env}
\end{figure}

\subsection{Packaging Worker-Learner into Pod}
\label{pod}

FinRL-Podracer achieves effective and efficient allocation of cloud resources through a multi-level mapping, which follows the principle of \textit{decomposition}-and-\textit{encapsulation}. We employ a worker-learner decomposition \cite{espeholt2020seedrl, espeholt2018impala, hessel2021podracer} that splits a DRL training process (pod) into three components (containers):

\begin{itemize}
    \item \textbf{Worker (exploration)}: samples transitions through the actor-environment interactions.
    \item \textbf{Replay buffer}: stores transitions from a worker and feeds a batch of transitions to a learner.
    \item \textbf{Learner (exploitation)}: consumes transitions and trains the neural networks.
\end{itemize}

Each training process of an agent consists of the three types of components, which are packaged into a suite that is mapped into a GPU pod. In addition, We run those components separately where each component is mapped to a GPU container. Such a two-level mapping is natural since a GPU pod consists of multiple containers, while correspondingly a training process of an agent consists of different components.  

The above pod-container structure enables scalable allocation of GPU computing resources. We take advantage of a GPU cloud software stack and use the Kubernetes (K8S) software to scalably coordinate pods among severs. Consider a cloud with $10$ servers (i.e., $80$ A100 GPUs), we encapsulate a package of components into a pod, replicate it $80$ times, and send them to a K8S server. Then, K8S distributes these $80$ pod replicas to computing nodes that carry out the training process. The pod replication reflects strong parallelism, and it is highly scalable since a GPU cloud can support a large number of pods. 


\subsection{High Performance Training Layer}

The optimization of each component is critical to the overall performance. We describe the hardware-oriented optimizations of components, including parallelism encapsulation, GPU acceleration, efficient parameter fusion and storage optimization.

\textbf{Market simulation with GPU-acceleration}. The market simulation is both computing- and communication-intensive. We propose a batch mode to perform massively parallel simulations, which maximizes the hardware utilization (either CPUs or GPUs). We instantiate multiple independent sub-environments in a batched environment, and a batched environment is exposed to a rollout worker that takes a batch of actions and returns a batch of transitions.

Fig. \ref{fig_env} a) illustrates a GPU-accelerated environment. Environments of financial tasks are highly suitable to GPUs because financial simulations involve "simple" arithmetics, where a GPU with thousands of cores has the natural advantages of matrix computations and parallelism. Then, financial environments written in CUDA can speed up the simulation. The GPU-accelerated environment also effectively reduces the communication overhead by bypassing CPUs, as supported by a GPU cloud \cite{NVIDIA_SupPod2020}. The output transitions are stored as a tensor in GPU memory, which can be directly fetched by learners, avoiding the data transfer between CPU and GPU back and forth.

Fig. \ref{fig_env} b) presents an environment on CPUs. There are some financial simulations with frequent CPU usage (addressing trading constraints), making it inefficient to compute on GPUs. In our experiments, some environments run much slower on GPUs than CPUs. Thus, we simulate those environments on CPUs.


\textbf{Replay buffer on GPU}. We allocate the replay buffer on the contiguous memory of GPUs, which increases the addressing speed and bypasses CPUs for faster data transfer. As the worker and learner are co-located on GPUs, we store all transitions as tensors on the contiguous memory of GPUs. Since the collected transitions are packed together, the addressing speed increases dramatically well when a learner randomly samples a batch of transitions to update network parameters.

\textbf{Learner with optimizations}. To better support the ensemble training in the evolution layer, we propose a novel and effective way for learners of each pod to communicate, i.e., sending the network parameters rather than the gradients. Most existing libraries \cite{liang2017rllib, stable-baselines, liu2020finrl, liu2021finrl} send the gradients of learners by following a traditional synchronization approach on supervised learning. Such an approach is inefficient for DRL algorithms since the learner will update the neural network hundreds of times within each training epoch, namely it needs to send gradients hundreds of times. By taking advantage of the soft update \cite{Lillicrap2016ContinuousCW}, we send the model parameters rather than the gradients. The parameter of the models is amenable to communication because model size in DRL is not comparable to that in other deep learning fields. Here, communication happens once at the end of each epoch, which is a significantly lower frequency of communication.

\begin{table}[t]
    \centering
    \begin{tabular}{ l l }
    \hline
    \textbf{Hyper-parameters} & \textbf{Value} \\
        \hline
        Total \#GPUs & 80\\
        \#Agent ($N$) & 10\\
        \#Pods per agent ($K$) & 8\\
        Optimizer & Adam\\ 
        Learning rate & $2^{-14}$\\ 
        Discount factor & $\gamma = 0.99$\\
        Total steps & $2^{20}$\\
        Batch size & $2^{10}$\\
        Repeat times & $2^{3}$\\
        Replay buffer Size & $2^{12}$\\
        Ratio clip (PPO) & 0.25\\
        Lambda entropy (PPO) & 0.02\\
        Evaluation interval (second) & $64$\\
        \hline
    \end{tabular}
    \caption{Hyper-parameter settings in our experiments.}
    \label{hyper}
    \vspace{-0.2in}
\end{table}

\section{Performance Evaluation}
\label{evaluation}
We describe the GPU cloud platform, the performance metrics and compared methods, and then evaluate the performance of FinRL-Podracer for a stock trend prediction task.

\subsection{GPU Cloud Platform}

All experiments were executed on NVIDIA DGX-2 servers \cite{choquette2021nvidia} in an NVIDIA DGX SuperPOD platform \cite{NVIDIA_SupPod2020}, a cloud-native supercomputer. We use 256 CPU cores of Dual AMD Rome 7742 running at 2.25GHz for each experiment. An NVIDIA DGX-2 server has $8$ A100 GPUs and $320$ GB GPU memory \cite{choquette2021nvidia}.

\begin{figure*}[!htb]
   \begin{minipage}{0.48\textwidth}
     \centering
     \includegraphics[width=3.3in]{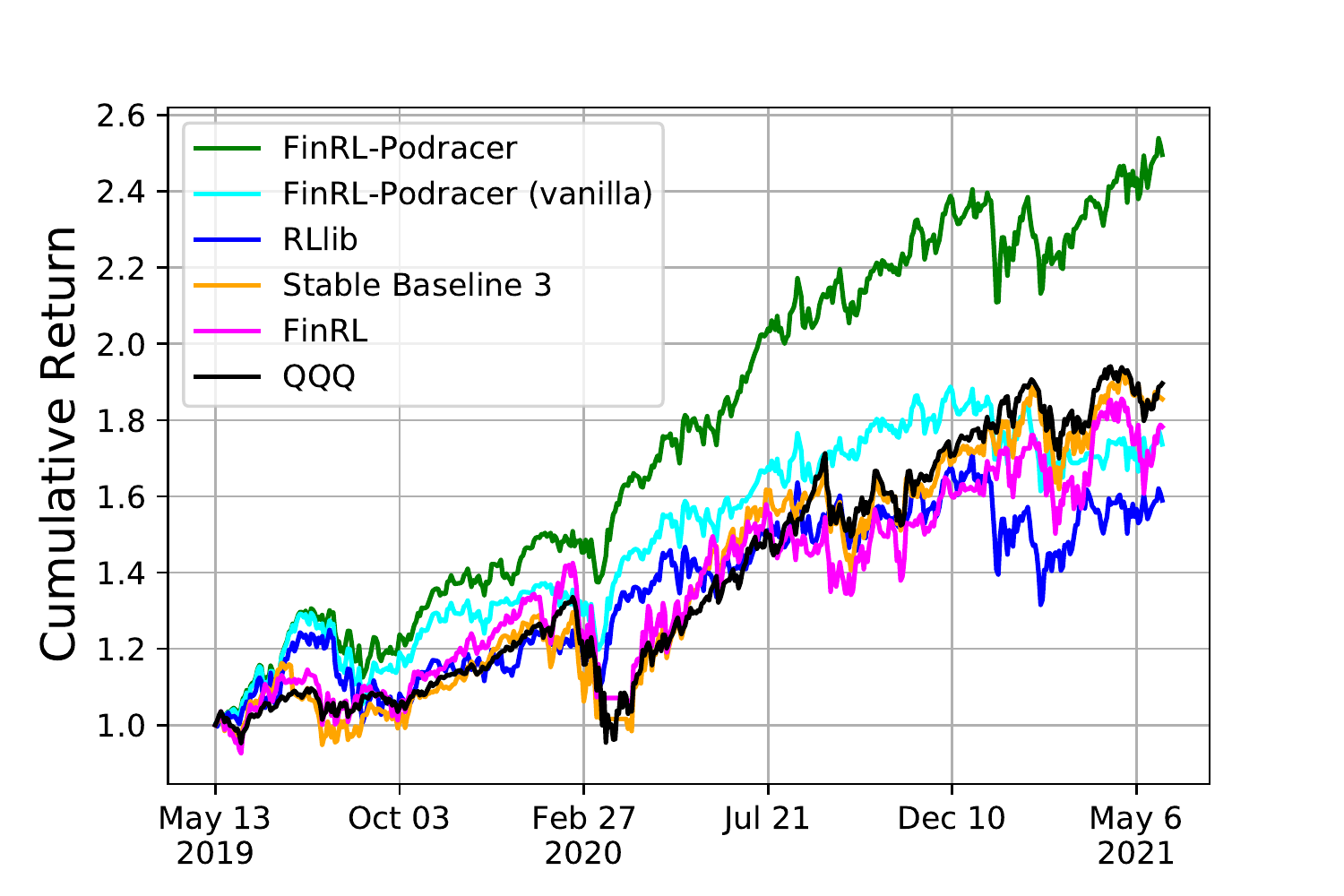}
    \caption{Cumulative returns on daily dataset during 05/13/2019 to 05/26/2021. Initial capital $\$1, 000, 000$, transaction cost percentage $0.2\%$, and Invesco QQQ ETF is a market benchmark.}
    \label{fig_cumulative_return_daily}
   \end{minipage}\hfill
   \begin{minipage}{0.48\textwidth}
     \centering
     \includegraphics[width=3.3in]{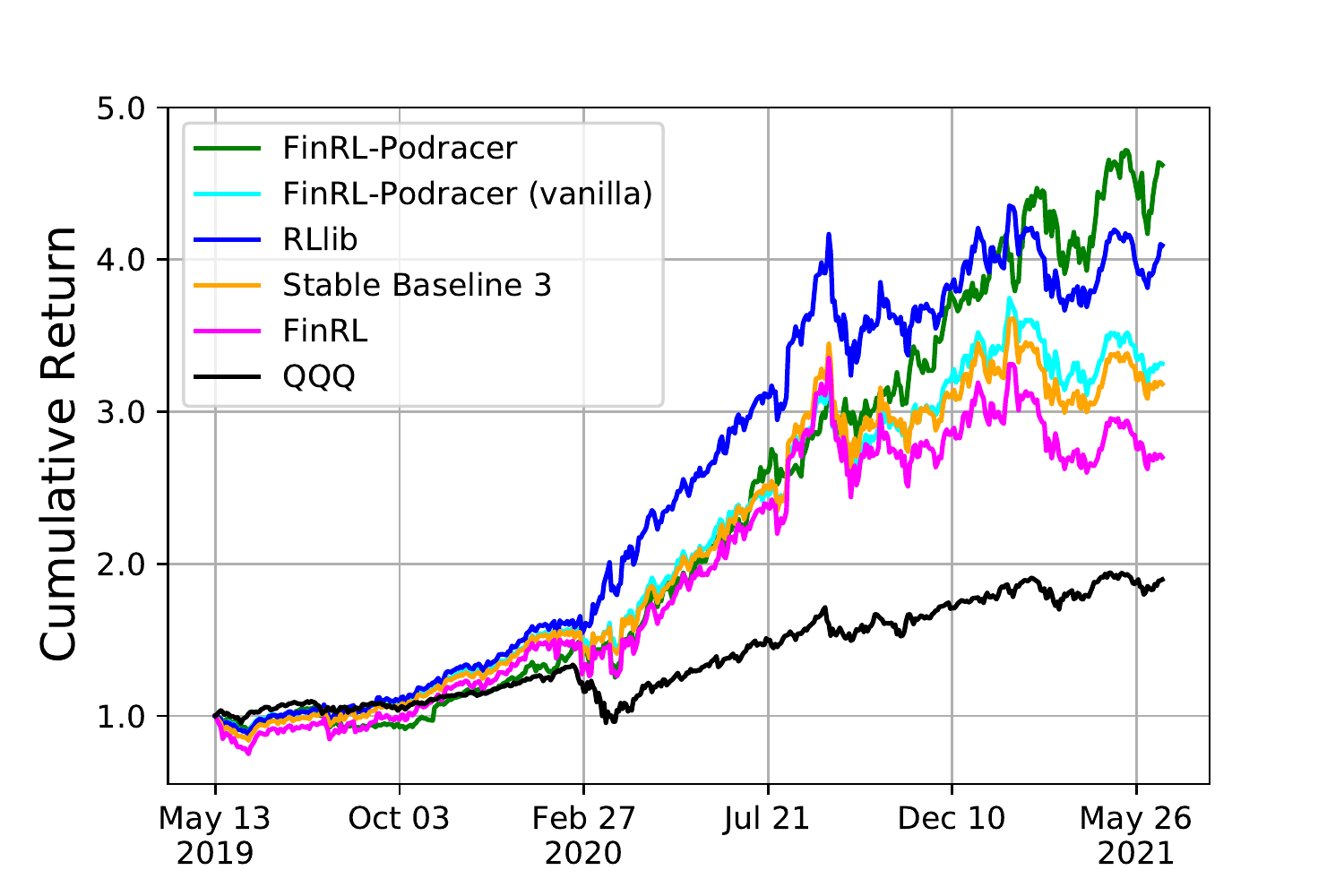}
    \caption{Cumulative returns on minute dataset during 05/13/2019 to 05/26/2021. Initial capital $\$1, 000, 000$, transaction cost percentage $0.2\%$, Invesco QQQ ETF is a market benchmark.}
    \label{fig_cumulative_return_minute}
   \end{minipage}
\end{figure*}

\begin{table*}
		\centering
		\begin{tabular}{|l|c|c|c|c|c|}\hline
			 & Cumul. return & Annual return & Annual volatility & Max drawdown & Sharpe ratio \\
			\hline
			FinRL-Podracer (\textbf{Ours}) &
			\textcolor{red}{\textbf{149.553\%}}/\textcolor{blue}{\textbf{362.408\%}}& \textcolor{red}{\textbf{56.431\%}}/\textcolor{blue}{\textbf{111.549\%}} &
			\textcolor{red}{\textbf{22.331\%}}/\textcolor{blue}{33.427\%} & \textcolor{red}{\textbf{-13.834\%}}/\textcolor{blue}{\textbf{-15.874\%}}& \textcolor{red}{\textbf{2.12}}/\textcolor{blue}{\textbf{2.42}}\\
			FinRL-Podracer (vanilla) &
			\textcolor{red}{73.546\%}/\textcolor{blue}{231.747\%} & \textcolor{red}{30.964\%}/\textcolor{blue}{79.821\%} &
			\textcolor{red}{23.561\%}/\textcolor{blue}{31.024\%} & \textcolor{red}{-18.428\%}/\textcolor{blue}{-21.002\%} & \textcolor{red}{1.27}/\textcolor{blue}{2.05}\\
			RLlib \cite{liang2017rllib} &
			\textcolor{red}{58.926\%}/\textcolor{blue}{309.54\%} & \textcolor{red}{25.444\%}/\textcolor{blue}{99.347\%} &
			\textcolor{red}{30.009\%}/\textcolor{blue}{31.893\%}  & \textcolor{red}{-23.248\%}/\textcolor{blue}{-22.292\%}& \textcolor{red}{0.91}/\textcolor{blue}{2.33}\\
			Stable Baseline3 \cite{stable-baselines} &
			\textcolor{red}{85.539\%}/\textcolor{blue}{218.531\%} & \textcolor{red}{35.316\%}/\textcolor{blue}{76.28\%} & \textcolor{red}{31.592\%}/\textcolor{blue}{34.595\%}  & \textcolor{red}{-24.056\%}/\textcolor{blue}{-23.75\%}& \textcolor{red}{1.12}/\textcolor{blue}{1.82}\\
			FinRL \cite{liu2020finrl} &
			\textcolor{red}{78.255\%}/\textcolor{blue}{169.975\%} & \textcolor{red}{32.691\%}/\textcolor{blue}{62.576\%} &
			\textcolor{red}{37.641\%}/\textcolor{blue}{42.908\%}  & \textcolor{red}{-26.774\%}/\textcolor{blue}{-27.267\%}& \textcolor{red}{0.94}/\textcolor{blue}{1.35}\\
			Invesco QQQ ETF & 89.614\% & 36.763\% & 28.256\% & -28.559\% & 1.25 \\
			\hline
		\end{tabular}
		\caption{Performance of stock trading on NASDAQ-100 constituent stocks with daily (\textcolor{red}{red}) and minute-level (\textcolor{blue}{blue}) data.}
		\label{tab:all}
		\vspace{-0.15in}
\end{table*}

\subsection{Performance Metrics}
\label{metrics}
We evaluate the trading performance and training efficiency of the FinRL-Podracer for a stock trend prediction task. 

\textbf{Data pre-processing}. We select the NASDAQ-100 constituent stocks as our stock pool, accessed at 05/13/2019 (the starting time of our testing period), and use the datasets with two time granularities: minute-level and daily. The daily dataset is directly downloaded from Yahoo!Finance, while the minute-level dataset is first downloaded as raw data from the Compustat database through the Wharton Research Data Services (WRDS) \cite{wrds} and then pre-processed to an open-high-low-close-volume (OHLCV) format. We split the datasets into training period and backtesting period: the daily data from 01/01/2009 to 05/12/2019 for training; the minute-level data from 01/01/2016 to 05/12/2019 for training; For both datasets, we backtest on the same period from 05/13/2019 to 05/26/2021.

\textbf{Evaluation metrics}. Six common metrics are used to evaluate the experimental results:
\begin{itemize}[leftmargin=*]
    \item \textbf{Cumulative return}: subtracting the initial value from the final portfolio value, then dividing by the initial value. 
    \item \textbf{Annual return and volatility}: geometric average return in a yearly sense, and the corresponding deviation.
    \item \textbf{Sharpe ratio}: the average return earned in excess of the risk-free rate per unit of volatility.
    \item \textbf{Max drawdown}: the maximum observed loss from a historical peak to a trough of a portfolio, before a new peak is achieved. Maximum drawdown is an indicator of downside risk over a time period.
    \item \textbf{Cumlative return vs. training time}: the cumulative return during the testing period, achieved by an agent trained within a certain amount of time.
\end{itemize}

\textbf{Compared methods}. For trading performance evaluation, we compare FinRL-Podracer and vanilla FinRL-Podracer (without agent evolution) with FinRL \cite{liu2020finrl, liu2021finrl}, RLlib \cite{liang2017rllib}, Stable Baseline3 \cite{stablebaseline3}, and NASDAQ Composite/Invesco QQQ ETF. We use Proximal Policy Optimization (PPO) \cite{Schulman2017ProximalPO} as the DRL algorithm in the reported results and fine-tune each library to maximize its performance. Each library is allowed to use up to 80 GPUs. 

For training efficiency evaluation, the experiments are conducted on multiple GPUs. We compare with RLlib \cite{liang2017rllib} since it has high performance on distributed infrastructure. However, both FinRL \cite{liu2020finrl} and Stable Baseline 3 \cite{stablebaseline3} do not support the training on multiple GPUs, thus we do not compare with them. We keep hyper-parameters and computing resources the same to guarantee fair comparisons, and a general hyper-parameter setting is given in Table \ref{hyper}. 

\begin{figure*}[!htb]
   \begin{minipage}{0.48\textwidth}
     \centering
     \includegraphics[width=3.3in]{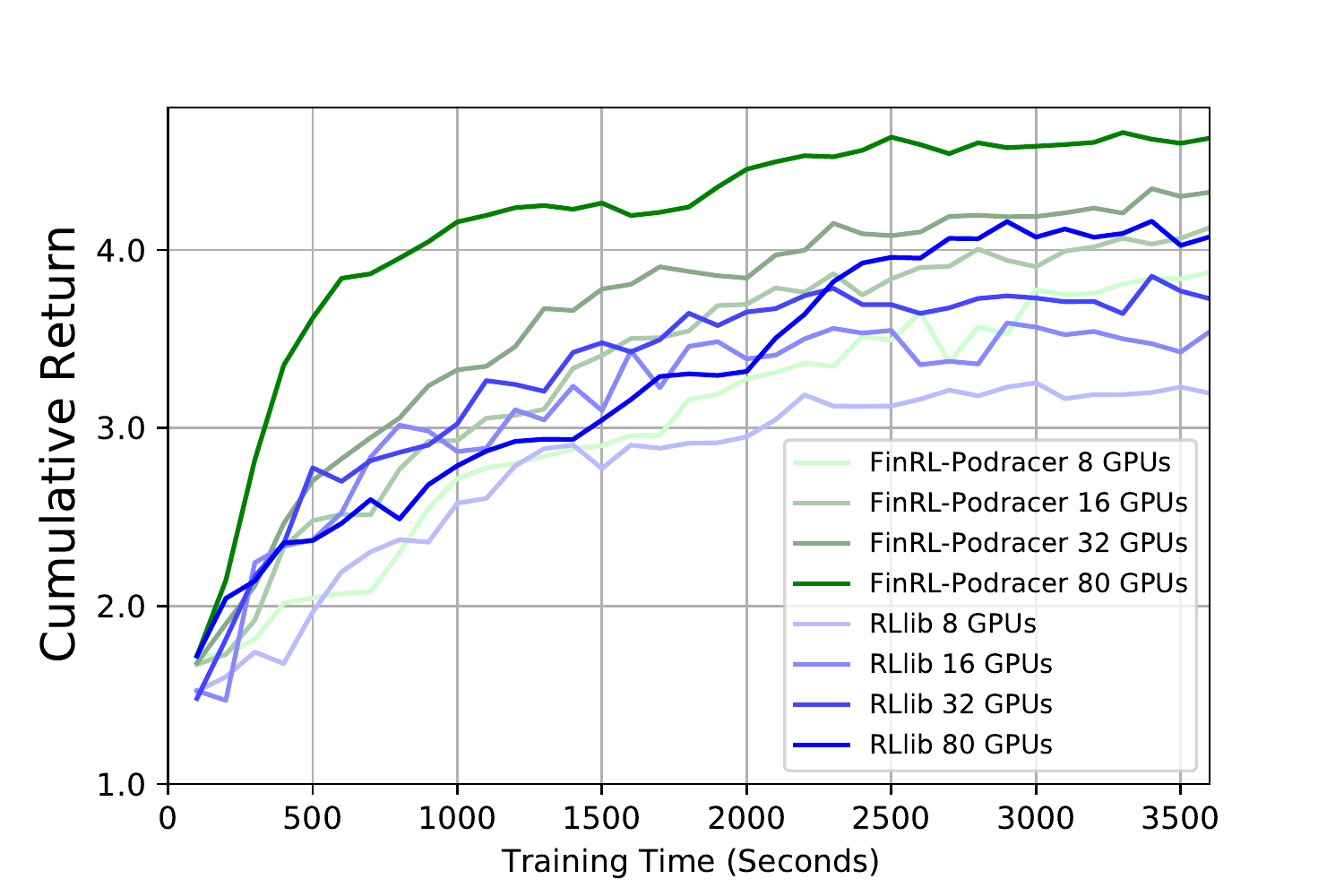}
    \caption{Generalization performance on backtesting dataset, using the model snapshots of FinRL-Podracer and RLlib \cite{liang2017rllib} at different training time (wall clock time).}
    \label{fig_efficiency}
   \end{minipage}\hfill
   \begin{minipage}{0.48\textwidth}
     \centering
     \includegraphics[width=3.3in]{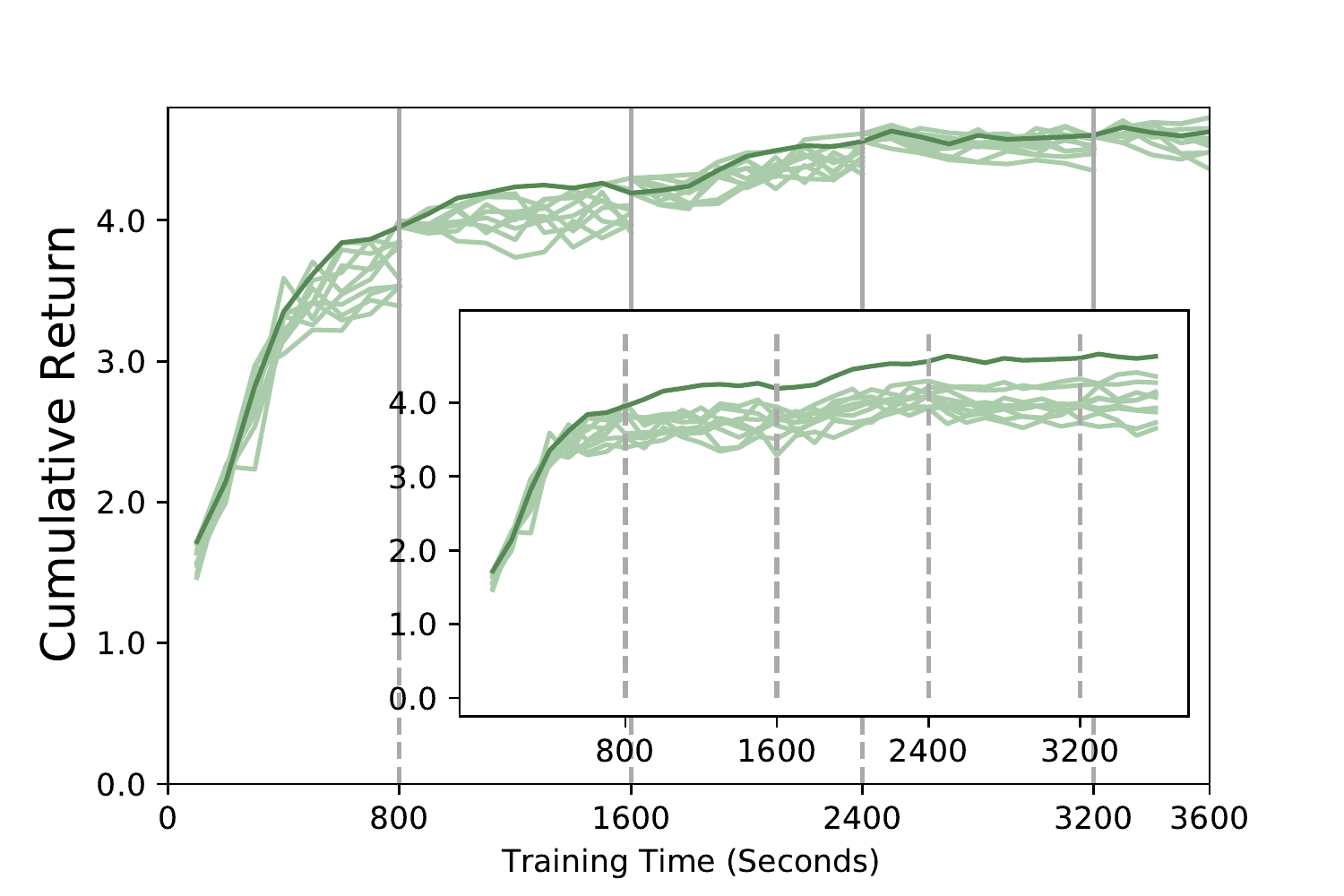}
    \caption{Generalization performance of a model along the agent evolution in the training process of FinRL-Podracer.}
    \label{fig_learning_curve}
   \end{minipage}
   \vspace{-0.2in}
\end{figure*}

\subsection{Trading Performance}

We backtest the trading performance from 05/13/2019 to 05/26/2021 on both daily and minute-level datasets. From Fig. \ref{fig_cumulative_return_daily} and Fig. \ref{fig_cumulative_return_minute}, all DRL agents are able to achieve a better or equal performance than the market in cumulative return, which demonstrates the profit potentials of DRL-driven trading strategies. Comparing Fig. \ref{fig_cumulative_return_daily} with Fig. \ref{fig_cumulative_return_minute}, we observe that all methods have a much better performance on the minute-level dataset than that on the daily dataset. The trading performance of most agents is almost the same as that of the market on daily dataset, however, all agents significantly outperform the market if they have a larger dataset to explore. With a higher granularity data, the Sharpe ratios are also lifted up to a new level. From Table \ref{tab:all}, agents achieve Sharpe ratios of 2.42, 2.05, 2.33, 1.82, 1.35 on the minute-level dataset, which are 0.3, 0.78, 1.42, 0.7, and 0.41 higher than those on the daily dataset. Therefore, we conclude that the capability to process large-scale financial data is critical for the development of a profitable DRL-driven trading strategy since the agent can better capture the volatility and dynamics of the market.

From Table \ref{tab:all}, Fig. \ref{fig_cumulative_return_daily}, and Fig. \ref{fig_cumulative_return_minute}, we also observe that our FinRL-Podracer outperforms other baselines on both datasets, in terms of expected return, stability, and Sharpe ratio. As can be seen from Table \ref{tab:all}, our FinRL-Podracer achieves the highest cumulative returns of 149.533\% and 362.408\%, annual returns of 56.431\% and 111.549\%, and Sharpe ratios of 2.12 and 2.42, which are much higher than the others. Furthermore, FinRL-Podracer also shows an outstanding stability during the backtesting: it achieves Max drawdown -13.834\% and -15.874\%, which is much lower than other methods. Consider the vanilla FinRL-Podracer as a direct comparison, we find that vanilla FinRL-Podracer has a similar trading performance with other baseline frameworks, which is in consistent with our expectation since the settings are the same. Such a performance improvement of FinRL-Podracer over vanilla FinRL-Podracer demonstrates the effectiveness of the generational evolution mechanism, as further verified by Fig. \ref{fig_learning_curve}.

\subsection{Training Efficiency}

We compare the training efficiency of FinRL-Podracer with RLlib \cite{liang2017rllib} on a varying number of A100 GPUs, i.e., 8, 16, 32, and 80. We store the model snapshots at different training time, say every $100$ seconds, then later we use each snapshot model to perform inference on the backtesting dataset and obtain the generalization performance, namely, the cumulative return. 

In Fig. \ref{fig_efficiency}, as the number of GPUs increases, both FinRL-Podracer and RLlib achieve a higher cumulative return with the same training time (wall-clock time). FinRL-Podracer with $80$ GPUs has a much steeper generalization curve than others, e.g., it can achieve a cumulative return of $4.0$ at $800$s, which means it learns in a much faster speed. However, FinRL-Podracer with $32$ GPUs and $16$ GPUs need $2,200$s and $3,200$s to achieve the same cumulative return, respectively. The generalization curves of RLlib with different numbers of GPUs are relatively similar, and we do not observe much speed-up. For example, FinRL-Podracer needs approximately $300$s to achieve a cumulative return of $3.5$, however, RLlib needs $2,200$s to achieve the same cumulative return. FinRL-Podracer is $3\times \sim 7\times$ faster than RLlib. 

\textbf{It is counter-intuitive that the increase of GPU resources not only makes FinRL-Podracer have a fast training, but also improves the trading performance over RLlib \cite{liang2017rllib}}. We know from Fig. \ref{fig_cumulative_return_daily} and Fig. \ref{fig_cumulative_return_minute} that the generational evolution mechanism promotes the trading performance of FinRL-Podracer, therefore, we empirically investigate the agent evolution process. Fig. \ref{fig_learning_curve} explicitly demonstrates an evolution of $N = 10$ agents, where the selector chooses the best model to train in the next generation every $800$s. 
The inner figure of Fig. \ref{fig_learning_curve} depicts the generalization curves of the ten agents in the first generation (without using the agent evolution mechanism). The curve with the evolution mechanism (the thick green curve) is substantially higher than the other ten curves. 


\section{Discussion and Conclusion}

In this paper, we have proposed a high-performance and scalable deep reinforcement learning framework, \textit{FinRL-Podracer}, to initiate a paradigm shift from conventional supervised learning approaches to \textit{RLOps in finance}. FinRL-Podracer provides a highly automated development pipeline of DRL-driven trading strategies on a GPU cloud, which aims to help finance researchers and quantitative traders overcome the steep learning curve and take advantage of supercomputing power from the cloud platforms.

FinRL-Podracer achieved promising performance on a cloud platform, mainly by following the two principles, \textit{the virtues of nested hierarchies} and \textit{getting smart from dumb things} \cite{kelly1994out}. For low-level training, FinRL-Podracer realizes nested hierarchies by empolying hardware-oriented optimizations, including parallelism encapsulation, GPU acceleration, and storage optimizations. As a high level scheduling, FinRL-Podracer obtains a smart agent from hundreds of weak agents, which is the essence of ensemble methods, by employing a generational evolution mechanism. We further investigate the evolution and training layers in a followup work \cite{liu2021podracer} for a cloud-native solution. We believe that ensemble multiple weak agents is preferable to aiming to train one strong agent. Thus we propose a new orchestration mechanism, a tournament-based ensemble training method \cite{liu2021podracer} with asynchronous parallelism, which involves relatively low communication overhead.  Also, we observe the great potential of massively parallel simulation, which lifts the exploration capability up into a potentially new dimension.


FinRL-Podracer is our first step from building a standard DRL pipeline of financial tasks to using DRL agents to understand the dynamics of the markets. We believe that FinRL-Podracer is critical for the ecosystem of the FinRL community \cite{liu2020finrl, liu2021finrl} because it offers opportunities for many future directions. First, FinRL-Podracer provides a way to take advantage of large-scale financial data. It is possible to allow DRL agents to work in second or microsecond level and cover all stocks in the market, which is meaningful for the exploration and understanding of the dynamics of the market. Moreover, training on the cloud makes DRL agents adapt to much more complex financial simulations and neural networks, thus achieving wider DRL applications to various financial tasks, e.g., portfolio allocation, fraud detection, DRL-driven insights for yield improvement and optimization. Furthermore, the low-level optimizations in FinRL-Podracer could be also useful for the future development of financial simulators, such as using GPU-accelerated techniques to reduce latency.




\section*{Acknowledgement}


This research used computational resources of the GPU cloud platform \cite{NVIDIA_SupPod2020} provided by the IDEA Research institute.

\bibliographystyle{ACM-Reference-Format}
\bibliography{main}

\end{document}